\documentclass[
aps,nofootinbib,
showpacs,showkeys,preprint
tightenlines,preprintnumbers,] {revtex4}

\usepackage{epsf,epsfig,subfigure,graphicx,amsmath,amssymb 
}
\usepackage{color}
\newcommand{\dis}[1]{\begin{equation}\begin{split}#1\end{split}\end{equation}}
\newcommand{\ie}{{\it i.e.~}}
\newcommand{\etal}{{\it et al.\,}}

\newcommand{\Qem}{Q_{\rm em}}

\newcommand{\tev}{\,\textrm{TeV}}
\newcommand{\gev}{\,\textrm{GeV}}

\newcommand{\Mg}{{M_{\rm GUT}}}

\def\sw0{{$\sin^2\theta_W^0$}}

\newcommand{\Z}{{\bf Z}}

\def\E6{{\rm E_6}}

\def\EE8{{\rm E_8\times E_8'}}

\begin{document}

\draft

\title{\Large\bf Is an axizilla possible for di-photon resonance?}

\author{ Jihn E.  Kim}
\address
{Department of Physics, Kyung Hee University, 26 Gyungheedaero, Dongdaemun-Gu, Seoul 02447, Republic of Korea, and\\
Center for Axion and Precision Physics Research (IBS),
  291 Daehakro, Yuseong-Gu, Daejeon 34141, Republic of Korea
}

\begin{abstract} 
 Heavy  axion-like particles, called  axizillas, are simple extensions of the standard model(SM). An axizilla is required not to couple to the quarks, leptons, and Brout-Englert-Higgs doublets of the SM, but couple to the gauge anomalies of the $W^\pm, Z$ and photon. It is possible to have its branching ratios(BRs) to two photons greater than 10\,\%  and to two $Z$'s less than 10\,\%. To have a (production cross section)$\cdot $(BR  to di-photons) at a $10^{-38}\,\textrm{cm}^2$ level, a TeV scale heavy quark $Q$ is required for the gluon--quark fusion process. The decay of $Q$ to axizilla plus quark, and the subsequent decay of the axizilla to two photons can be fitted at the required level of $10^{-38}\,\textrm{cm}^2$.
 
\end{abstract}
\maketitle


\section{Introduction}\label{sec:Introduction}

The recent report on possible di-photon events at 750 GeV from the LHC Run-II experiments \cite{ATLASCMS,LHCrun2,diPhoton,CERNth}  triggered a lot of theoretical interest on this issue. The requirement to explain the rate is to require the production rate,  $\sigma_{\rm production}\cdot $(branching ratio(BR) to di-photons), of order $10^{-38}\,\textrm{cm}^2$ with the LHC parton distribution at 13 TeV  energy. Any model for a diphoton resonance of mass 750 GeV decaying to two photons is better not to give such di-phtons at the previous LHC Run-I at 8 TeV.  

This invites to search for, ``Which particle is most economically introduced beyond the standard model(SM)?'' A phenomenology on this is summarized in Ref. \cite{CERNth} and the recent papers are listed in \cite{Das15,More}. Here, we search for a theory motivated particle. The well-known examples are axions \cite{KimRMP10}, majorons \cite{CMPmajoron},  ALPs \cite{ALPs}, and quintessencial pseudoscalars or ultralight axions \cite{KimNilles14}, which are much lighter than electron. We argue that not only these very light pseudoscalars but also TeV scale pseudoscalars are theoretically prospective. Since pseudoscalars are pseudo-Goldstone bosons of some spontaneously broken axial symmetry, we call these heavy pseudoscalars {\it axizillas}. In this paper, we investigate a possibility of an axizilla for the di-photon resonance of the LHC Run II. 

It is known that string theory does not allow any global symmetry below the compactification scale, except the model-independent axion \cite{Witten84,Wen86}. Also from the topological obstruction of global symmetries from gravity \cite{Coleman88}, it has been argued that a serious fine-tuning problem is present in axion physics \cite{Barr92}. In string compactification with the anomalous U(1) gauge symmetry, a global Peccei-Quinn(PQ) symmetry can survive down to an intermediate scale \cite{KimPLB88}. The resulting invisible axion is from a phase field of matter scalars instead from the anti-symmetric tensor field $B_{\mu\nu}$ \cite{KimPRL13}. Except from the anomalous U(1) gauge symmetry, any global symmetry must be approximate. For example, it has been shown numerically  that there exist compactification models suppressing   the explicit PQ symmetry breaking terms at some level \cite{IWKimJHEP07}.
 Generically, however, any global symmetry in consideration must be broken at some leading scale below the compactification scale. Nevertheless some discrete symmetries can be allowed without the gravity obstruction problem \cite{Krauss89}. 
 
\begin{figure}[!t]
\begin{center}
\includegraphics[width=0.35\linewidth]{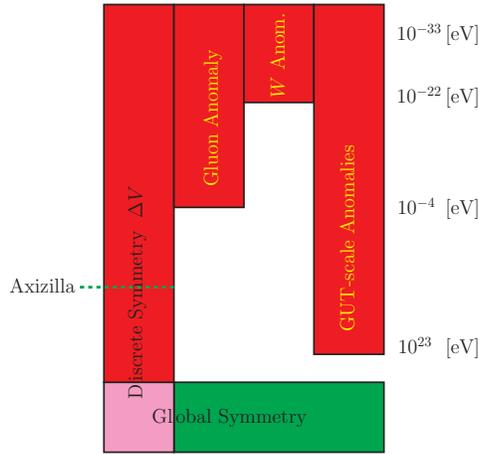}
\end{center}
\caption{A cartoon of classifying symmetries at low energy. The terms in the vertical column are allowed by discrete symmetries in string compactification. 
The terms in the superpotential belongs to the most left column. The other three columns show anomalous terms.
If one consideres a few leading effective terms, \ie corresponding to the lavender square, the terms there define an effective global symmetry. However, this global symmetry is broken by the terms in the red boxes.  Some mass scales needed in cosmology are also shown with the axion mass scales \cite{Quint,ULA,Kim79,GUTa}.} \label{fig:effGlobal}
\end{figure}

 In Fig. \ref{fig:effGlobal}, we show the superpotential terms allowed by the discrete symmetry in the most left vertical column. 
 If we consider a few lowest order terms, \ie those inside the lavender square, there might appear a global symmetry. The terms allowed by this global symmetry are shown in the horizontal bar including those in the green box. Thus, this global symmetry is broken by the terms in the most left red boxes in the vertical column. In addition, we have shown also the non-Abelian anomaly terms which also break the global symmetry. 
 If the PQ type global symmetry is respected by the superpotential, we neglect the most left column. In this case, the $\theta$ angle of the non-Abelian group vacuum chooses $\theta=0$ as the minimum, which is used in
 extremely light axions, quintessential axion \cite{Quint}, ultra light axion \cite{ULA}, QCD axion \cite{Kim79}, and axionic inflation \cite{GUTa}. The respective axion mass scales are shown at the far right.
 On the left-hand side (LHS), the mass scale of axizilla is shown. The breaking scale is not known in any known non-Abelian gauge symmetry, and its mass derives from the global symmetry breaking potential, $\Delta V$.
Let the global symmetry be U(1)$_\Sigma$. Suppose $S$ carries the discrete quantum number but is neutral under U(1)$_\Sigma$ and $\sigma$ carries both the discrete and U(1)$_\Sigma$ quantum numbers, 
 \dis{
 &S,~~\textrm{with a GUT scale VEV~}\Mg,\textrm{ but not breaking U(1)}_{\rm global},\\
&\sigma,~~\textrm{with a VEV }f/\sqrt2, \textrm{  breaking U(1)}_{\rm global}. 
 }
For some discrete symmetry $\Z_{N}$, assign the global quantum numbers $\Sigma$ as those of $\Z_N$ with $-N< \Sigma<N$. For a discrete symmetry $\Z_{4}$  for example, let the discrete quantum numbers of $\sigma$ be 1. Then, the following global symmetry breaking term, allowed by $\Z_4$, is present 
\dis{
\Delta V\sim \sigma^4 +{\rm h.c.} \to   f^4 \cos\left( 4\frac{P}{f}\right) \nonumber
}
where $|\sigma|=f/\sqrt2$. Without $S$ fields, if $f$ is of order TeV scale, which can be determined by the U(1)$_\Sigma$ preserving terms, then the pseudoscalar mass $m_P$ is at the TeV scale. This is an axizilla. With $S$ fields included, more complicated discrete symmetries can produce TeV scale axizillas. A cosmological effect of $\Z_N$ symmetry is the appearance of domain walls \cite{Okun74} which however is difficult to be observed after inflation.

\section{Axizillas with discrete symmetry $\Z_N$}

With a discrete  symmetry, we try to introduce a global anomaly but without a color anomaly. If it has a color anomaly, it is necessarily related to the QCD axion and the symmetry breaking scale must be larger than $10^{10\,}\gev$. As the simplest example, let us introduce a vectorlike doublet with the hypercharge $Y=-\frac12$,
 \dis{
\ell_L= \begin{pmatrix}
  N \\   L \end{pmatrix}_L , \quad 
 \ell_R =\begin{pmatrix}
  N \\   L \end{pmatrix}_R .
  \label{eq:Yukawa2}
 } 
In the SM, the pseudoscalar ($P$) coupling to color singlet gauge bosons  are
 \dis{
{\cal L}= \frac{ P}{f}\,\frac{g_2^2}{32\,\pi^2}W^{a}_{\mu\nu}\tilde{W}^{{b}\,\mu\nu}\left(\textrm{Tr\,}T_aT_b\right)+ \frac{ P}{f}\,\frac{g^{\prime \,2}}{32\,\pi^2}Y_{\mu\nu}\tilde{Y}^{\mu\nu}\left(\textrm{Tr\,}Y_L^2+
\textrm{Tr\,}Y_R^2\right)\label{eq:Model1}
 }
 where $W^{a}_{\mu\nu}$ is the non-Abelian field strength of SU(2) gauge fields $A^a_\mu$ and  $Y_{\mu\nu}$ is the field strength of U(1)$'$ gauge fields $Y_\mu$. For a vectorlike fundamental representation in SU($N$), like a heavy quark axion model \cite{Kim79} or Eq. (\ref{eq:Yukawa2}), $\textrm{Tr\,}T_aT_b=\frac12 \delta_{ab}$.
 Thus, Eq. (\ref{eq:Model1}) becomes
 \dis{
{\cal L} &= \frac{ P}{f}\,\frac{g_2^2}{32\,\pi^2}W^{a}_{\mu\nu}\tilde{W}^{{a}\,\mu\nu} + \frac{P}{f}\,\frac{g^{\prime \,2}}{32\,\pi^2}
Y_{\mu\nu}\tilde{Y}^{\mu\nu}\\[0.5em]
&=\frac{P}{32\,\pi^2\, f}\left(2g_2^2 W^{+}_{\mu\nu}\tilde{W}^{{-}\,\mu\nu} +2e^2 F^{\textrm{em}}_{\mu\nu}\tilde{F}^{\textrm{em}\,\mu\nu}+ g_2^2 (1/c_W^2-2 s_W^2)  Z_{\mu\nu}\tilde{Z}^{\mu\nu} +2\frac{eg_2}{c_W } F^{\textrm{em}}_{\mu\nu}\tilde{Z}^{\mu\nu}
\right)
 \label{eq:Model2}
 }
 where
 \dis{
 &W^3_\mu=\cos\theta_W Z_\mu +\sin\theta_W A_\mu,\\
 &Y_\mu=-\sin\theta_W Z_\mu +\cos\theta_W A_\mu ,\\
 &c_W=\cos\theta_W=\frac{g_2}{\sqrt{g_2^2+g^{\prime\,2}}},\\
 &s_W=\sin\theta_W=\frac{g'}{\sqrt{g_2^2+g^{\prime\,2}}}.
 }
The massless combination to the photon coupling is parametrized by $c_{P\gamma\gamma}$,
\dis{
{\cal L}_{P\gamma\gamma}= \frac{c_{P\gamma\gamma} P}{f}\,\frac{e^2}{32\,\pi^2}F^{\rm em}_{\mu\nu}\tilde{F}^{{\rm em}\,\mu\nu}
 }
where $c_{P\gamma\gamma}$ turns out to be 2.
\begin{table}[t!]
\begin{center}
\begin{tabular}{c|cccccc} \\[-1.2em] &$\ell_L$ &$\ell_R$ &   $\sigma$& $S$ &$E_L$&$E_R$  \\[0.2em]\hline &&& &&& \\[-1.2em] 
$\Z_{12}$ & ~$-\frac12$~ & ~$+\frac12$~ & ~$1$~    & ~$+4$~& ~$+\frac12$~&~$-\frac12$~ \\[0.2em]
$\Sigma$ & ~$-\frac12$~ & ~$+\frac12$~ & ~$1$~    & ~$+4$~& ~$+\frac12$~&~$-\frac12$~ \\[0.2em]\hline
 \end{tabular}
\end{center}
\caption{The U(1)$_\Sigma$ quantum numbers. $\Delta V$ can contain $\sigma^4 (S^*+S^2)$.}\label{tab:QuanNoQ}
\end{table}
Treating the $W$ and $Z$ bosons as massless, we estimate the branching ratios (BRs), to the decay modes to $W^+W^-, 2\gamma, 2Z,$ and $Z\gamma$ of Eq.  (\ref{eq:Yukawa2}), for which the BRs is shown in the first row of Table \ref{tab:BRs}, where we used $\sin^2\theta_W\simeq 0.23$. 
The effect of $\Z_N$ symmetry is to constrain possible interactions such that the leading term is suppressed by one power of $f$. Without the $\Z_N$ symmetry, some terms in the potential dominate this anomaly term \cite{Barr92}.

Let us now introduce $n_1$ pairs of $\Qem=-1$ vectorlike SU(2) singlet $E$ and $n_2$ pairs of (\ref{eq:Yukawa2}),
 \dis{
n_2\left\{ \ell_L= \begin{pmatrix}
  N \\   L \end{pmatrix}_L , ~
 \ell_R =\begin{pmatrix}
  N \\   L \end{pmatrix}_R\right\}, \quad n_1 \Big\{E_L,~ E_R\Big\}.
  \label{eq:Yukawa12}
 } 
Now, we have
  \dis{
{\cal L}_{P\rm-decay} &= n_2\frac{ P}{f}\,\frac{g_2^2}{32\,\pi^2}W^{a}_{\mu\nu}\tilde{W}^{{a}\,\mu\nu} - (2n_1-n_2)\frac{P}{f}\,\frac{g^{\prime \,2}}{32\,\pi^2}
Y_{\mu\nu}\tilde{Y}^{\mu\nu}\\[0.5em]
&=\frac{P}{32\,\pi^2\, f}\Big( g_2^2\,2n_2 W^{+}_{\mu\nu}\tilde{W}^{{-}\,\mu\nu} -2g_2^2 s_W^2(n_1-n_2) F^{\textrm{em}}_{\mu\nu}
\tilde{F}^{\textrm{em}\,\mu\nu}\\
&\quad+ \frac{g_2^2}{c_W^2} (n_2-2 s_W^2[n_1s_W^2+n_2c_W^2])  Z_{\mu\nu}\tilde{Z}^{\mu\nu} +2g_2^2\frac{s_W}{c_W}  \left[n_2 -2n_1 s_W^2 \right] F^{\textrm{em}}_{\mu\nu}\tilde{Z}^{\mu\nu} \Big).\label{eq:Model3}
 }
 In Table \ref{tab:BRs}, we present the BRs for several values of $n_2$ and $n_1$. Note that the BR to two photons can be made significantly larger than the BR to $2Z$. For example, $n_2=2$ and $n_1=5$ gives the LHC di-photons but two $Z$'s within the experimental error bound.
 
\begin{table}[t!]
\begin{center}
\begin{tabular}{cc|ccc  c}
 \hline &&&&& \\[-1.2em] $n_2$&$n_1$ &$W^+W^-$ &   $2\gamma$& $2Z$ &$Z\gamma$   \\[0.2em]\hline &&&& &  \\[-1.2em] 
$1$ & ~$0$~ & ~$0.73$~ & ~$0.005$~    & ~$0.10$~& ~$0.17$~ \\[0.2em]
$1$ & ~$5$~ & ~$0.58$~ & ~$0.25$~    & ~$\sim 0$~ & ~$0.17$~  \\[0.2em]
$1$ & ~$6$~ & ~$0.30$~ & ~$0.39$~    & ~$\sim 0$~ & ~$0.31$~  \\[0.2em]
$2$ & ~$4$~ & ~$0.58$~ & ~$0.03$~    & ~$0.07$~ & ~$0.32$~  \\[0.2em]
$2$ & ~$5$~ & ~$0.81$~ & ~$0.10$~    & ~$0.08$~ & ~$0.01$~  \\[0.2em]
$2$ & ~$6$~ & ~$0.72$~ & ~$0.15$~    & ~$0.06$~ & ~$0.07$~  \\[0.2em]
\hline
    \end{tabular}
\end{center}
\caption{Branching ratios of $P$.}\label{tab:BRs}
\end{table}

For the process, $q_i(p_1)+\bar{q}_j(p_2)\to W^\pm(k')+ P(p')$ with the intermediate $W^\mp$, the  cross section is estimated as
\dis{
\sigma &=\frac{1}{4\pi^2} \frac{1}{|{\bf v}_1 -{\bf v}_2|}\frac{1}{(2s_1+1)(2s_2+1)}\sum_{spins}\int \frac{d^3k'd^3 p'}{2^4 E^2E_kE_P} |T|^2\delta^{(4)}(p_1+p_2-k'-P')\\
&\simeq  \frac{\alpha_2^3}{512\pi}\frac{E_k^3E_P}{f^2\, E^4 } \left[ 1+\frac{E_k}{E_P}-\frac{m_P^2}{E_P^2}  \right] \approx O(10^{-8})\frac{1}{f^2}\approx O\left(10^{-41}\frac{1}{f_{\tev}^2}\right) ~ \textrm{cm}^2. \label{eq:Prodg2}
}
Taking $f\approx 10-100\,\tev$, the cross section is of order $10^{-44\,}\textrm{cm}^2$ which is O($10^{-6}$) too small, even before multiplying quark distribution functions, to interpret the LHC di-photons. If the gluon anomaly is introduced, then the production cross section is estimated to be
$(\alpha_3/\alpha_2)^3$ multiplied to the result Eq. (\ref{eq:Prodg2}). This improves somewhat but the gluon distribution at $x\sim 0.03$ is so small that production by the intermediate gluon line is not enough. From the gluon anomaly term, 
\dis{
{\cal L}_{Pgg}= \frac{c_{Pgg} P}{f}\,\frac{g_3^2}{32\,\pi^2}G^{a}_{\mu\nu}\tilde{G}^{a\,\mu\nu},\label{PggCoup}
 }
  the decay rate is given by
  \dis{
\Gamma \approx   1.14\times10^{-3}  \left( \frac{ m_{P}}{ \rm 0.75\,TeV}\right)^3   \, \frac{c_{Pgg}^2}{f_{\tev}^2}\,\gev.
}
Then, the gluon fusion process gives the cross section
\dis{
A^2\frac{1}{m_P\Gamma_P}\approx  1.3\times 10^{-5}
\frac{c_{Pgg}^2}{f^2} \,\frac{m_P }{\Gamma_P}\approx   5\times 10^{-39} 
\frac{c_{Pgg}^2}{f_{\tev}^2} \,\frac{m_{P} }{\Gamma_P}\,[{\rm cm}^2] \approx  3.3\times 10^{-33}  \left( \frac{\rm 0.75\,TeV}{ m_{P}}\right)^3\frac{c_{Pgg}^2}{f_{\tev}^2} \,\,[\rm cm^2] ,\label{eq:gluonbare}
}
where $\Gamma_P$ is the decay width of axizilla and $A^2=c_{Pgg}^2\alpha_3^2E^2/64\pi^2 f^2\approx 1.3\times 10^{-5}
c_{Pgg}^2m_P^2/f^2$. The gluon distribution function   is of order  $xg(x,Q^2)\approx 0.02$ at $x=0.1-0.001$ at $Q^2=2\gev^2$ \cite{CMSgluon14}. Using this number for both gluons at $x\simeq 0.01$, \ie at $130\,\gev$, the energy is not enough to produce the resonance. However to have enough energy, one may take $x=O(0.1)$  at least for one gluon. In this region, the product of distribution functions is of order $4\times 10^{-4}$. Multiplying this to (\ref{eq:gluonbare}), we obtain the $P$ production cross section of order
\dis{
\approx  1.3\times 10^{-37}  \left( \frac{ \rm 0.75\,TeV}{ m_{P}}\right)^3\frac{c_{Pgg}^2}{f_{\tev}^2} \,[\rm cm^2] .\label{eq:gluondist}
}
 If $P$ decays to two gluons, the BR to $W^+W^-$ is of order 10\,\% and  the BR to $2\gamma$ is of order 1\,\%. Thus, a rough estimate of $\sigma_{\rm production}\cdot $(branching ratio(BR) to di-photons) is of order $10^{-39}\,[{\rm cm}]^2/ f_{\tev}^2$. Since $f$ is expected to be of order $>O(10\,\tev)$, two gluon fusion seems not enough for the di-photon resonance.
 
 \section{Heavy quarks at TeV scale}
 
\begin{figure}[!t]
\begin{center}
\includegraphics[width=0.3\linewidth]{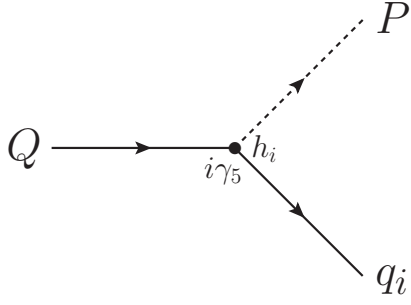}
\end{center}
\caption{The Feynman diagram for $Q\to P+q_i$.} \label{fig:Qdecay}
\end{figure}

Maybe, a heavier particle might have been produced such that its decay products include $P$. Since the data is compatible with the di-photon resonance production with little kinetic energy \cite{Unki}, the mass of this heavier particle may not be much heavier than 1\,TeV. Anticipating gluon--quark fusion process for the production of the heavier particle, let us introduce vectorlike heavy quark(s) $Q$'s. It must interact with gluon via the light quarks  and the heavy quark(s)  interaction
\dis{
{\cal L}=h_i \bar{q}_{iL}Q_RH+h.c.\to h_i\,P\, \bar{q}_{i}i\gamma_5Q
\label{eq:Qdecay}
}
where $h_i$ is the Yukawa coupling to the $i$-th light-quark doublet and
$Q_R$ is an SU(2)$_W$ doublet, and $H$ is an SU(2)$_W$ singlet.\footnote{The singlet $H$ should not be confused with the BEH doublet. To distinguish it from  the singlet notations $S, \sigma$ and $\sigma'$ of Table \ref{tab:QuanWithQ} and Eq. (1), we use $H$ as a singlet here since we need not introduce the notation of the BEH doublet in this paper. $Q$ is a vectorlike SU(2)$_W$ doublet, like (\ref{eq:Yukawa2}) with the additional degree of color.} Below a few TeV, the heavy component of $H$ is integrated out, leading to $P$ and $Q$ at the TeV scale.
The decay of $Q$ is depicted in Fig. \ref{fig:Qdecay}.
Then, using Eq. (\ref{eq:Qdecay}),  a magnetic moment type effective interaction is obtained,
 \dis{
 {\cal L}_{\rm Qqg}=\frac{{c}_i}{\Lambda}~G^{a\,\mu\nu}\bar{q}_i\frac{[\gamma_\mu,\gamma_\nu]}{2}F_aQ\label{eq:effGlQ}
 }
where $F^a$ is the generator of SU(3)$_c$, $G^{a\,\mu\nu}$ is the field strength of gluon field $G^a_\mu$ and $\Lambda$ is the effective mass scale. Due to the heavy quark coupling to $H$,  the axizilla coupling to the gluon anomaly is present. The partial decay rate to $q_i$ and gluon  is given by
\dis{
\Gamma_Q^{q_i}\simeq  \left(\frac{2\,c_i  m_Q}{\sqrt{\pi}\,\Lambda}\right)^2   \,m_Q.\label{eq:QtoGlue}
}   
The partial decay width due to (\ref{eq:Qdecay}) is
\dis{
\Gamma(Q\to q_i+P)=\frac{h_i^2}{4\pi}\,m_Q\left(1-\frac{M_P^2}{m_Q^2}\right).\label{eq:QtoP}
}
In some formulae below, to simplify the expression we assumed (\ref{eq:QtoP}) is sub-dominant compared to (\ref{eq:QtoGlue}).

Using Eq. (\ref{eq:effGlQ}),  we have the $u(p)_L+\textrm{gluon}(k)\to u(p')_L+\textrm{gluon}(k')$ with an intermediate heavy quark $Q(Q)$, whose amplitude is proportional to
 \dis{
 \bar{u}_L(p')\sigma^{\mu\nu}F^a G^a_{\mu\nu}(k')\frac{i}{Q\hskip -0.225cm\slash\,-m_Q}~F^bG^{b}_{\rho\sigma} (k) \sigma^{\rho\sigma}u_L(p)\label{eq:effSpinor}
 }
where $F^a$ is the generator of SU(3)$_c$, and $\sigma^{\mu\nu}=i[\gamma^\mu,\gamma^\nu]/2$. Note that the relevant light particles below a few TeV are heavy quarks and the axizilla $P$. The hypothetical global symmetry, called U(1)$_\Sigma$ is broken above a few TeV, and the scalar part of $H$ is assumed to be heavier than a few TeV.
So, if ${\cal L}_{\rm Qqg}$ in (\ref{eq:effGlQ}) is the only relevant  interaction, the cross section is estimated as,  for the head-on collisions in the proton + proton machine,
\dis{
\sigma(x_1,x_2)\simeq 
 \frac{ 3c_i^2}{ 2\Lambda^2} \frac{(m_{q_i}^2/E^2)  }{\pi  (x_1+x_2)   \left[(x_1x_2-\frac{m_Q^2}{4E^2})^2+\frac{\Gamma^2}{4E^2} \right]}\,  \ln\left(\frac{2}{1-\cos\theta_{\rm min}}\right),\label{eq:crossxx}
}
where $E$ is the beam energy, \ie $6.5\, \gev$ at Run-II, $m_{q_i}$ is   the light quark mass of species $q_i$, $x_1$ is the scaling variable of the incoming quark and $x_2$ is the scaling variable of the incoming gluon.

We attempt to solve the strong CP problem by the invisible axion. 
Since $Q$ carries color, we must worry about the QCD anomaly.
So,
if the QCD axion decay constant $f_a$ is near the TeV scale then the model is ruled out from the SN1987A bound \cite{SNbounds}. The gluon anomaly must be absent at the TeV scale.  This can be achieved by two axions such that the gluon anomaly is carried away by the invisible axion at a high energy scale \cite{KimRMP10} and the 750 GeV resonance does not carry the gluon anomaly
even though we introduced colored vectorlike doublets by $Q$.
So, we introduced another heavy quark $Q_2$ in addition to $Q_1$ such that two global symmetries,  the PQ symmetry
U(1)$_\Gamma$ and the 750 GeV resonance--related symmetry   U(1)$_\Sigma$, can be introduced.
In Table \ref{tab:QuanWithQ}, we presented this idea on the RHS of the double bar, where  $P$ is not coupling to the gluon anomaly and an invisible axion is introduced by the phase of $\sigma'$. For this to be an invisible axion, we forbid the terms of the form 
$(\sigma')^n$ for $n\le 8$. A natural solution of this kind from superstring compactification is through the anomalous U(1) \cite{KimPLB88}, in which case $\Delta V=0$, \ie  the most left red column of Fig. \ref{fig:effGlobal} is absent.\footnote{The anomalous U(1) from string compactification is a gauge symmetry, and hence there does not exist the wormhole problem \cite{KimPLB13}. The global symmetry below the anomalous U(1) scale is an excellent PQ symmetry from string compactification \cite{KimPRL13}.}
But, the model of  Table \ref{tab:QuanWithQ} must introduce Dirac masses $m_1$ and $m_2$ at the TeV scale  in $m_1 \overline{Q}_{2R} Q_{1L}+m_2 \overline{Q}_{1R}Q_{2L}+{\rm h.c}$. For the estimation of the cross section, we use the order given in Eq. (\ref{eq:crossxx}).

\begin{table}[t!]
\begin{center}
\begin{tabular}{c|cccc|cc||ccccccc}
 \hline &&&&&&&& \\[-1.2em] &$\ell_L$ &$\ell_R$ &   $\sigma$& $S$ &$E_L$&$E_R$ &$Q_{1L}$&$Q_{1R}$&$Q_{2L}$&$Q_{2R}$ &$Q'_L$&$Q'_R$&$\sigma'$  \\[0.2em]\hline &&&& && && \\[-1.2em] 
$\Z_{12}$ &$-\frac12$&$+\frac12$&~$1$~& ~$4$~&~$+\frac12$~&~$-\frac12$~& ~$+\frac12$~&~$-\frac12$~& ~$-\frac12$~&  ~$+\frac12$~&~$0$~&~$0$~&~$0$~\\[0.2em]
$\Sigma$ & ~$-\frac12$~ & ~$+\frac12$~ & ~$1$~& ~$4$~&~$+\frac12$~&  ~$-\frac12$~& ~$+\frac12$~&~$-\frac12$~&  ~$-\frac12$~&~$+\frac12$& ~$0$~&~$0$~&~$0$~\\[0.2em]
$\Z_{2}$ &~$0$& ~$0$& ~$0$&~$0$& ~$0$&  ~$0$~&~$0$~&~$0$~&~$0$~&~$0$~& ~$+\frac12$~&~$- \frac12$~&~$1$ \\[0.2em]
$\Gamma$ & ~$0$& ~$0$&~$0$~& ~$0$~& ~$0$&  ~$0$& ~$0$&  ~$0$& ~$0$&  ~$0$& ~$+\frac12$~&~$-\frac12$~&~$ 1$ 
\\[0.2em]\hline
 \end{tabular}
\end{center}
\caption{Quantum numbers of   U(1)$_\Sigma$ and U(1)$_\Gamma$ from $\Z_{12}\times\Z_2$.}\label{tab:QuanWithQ}
\end{table}

\begin{figure}[!t]
\begin{center}
\includegraphics[width=0.8\linewidth]{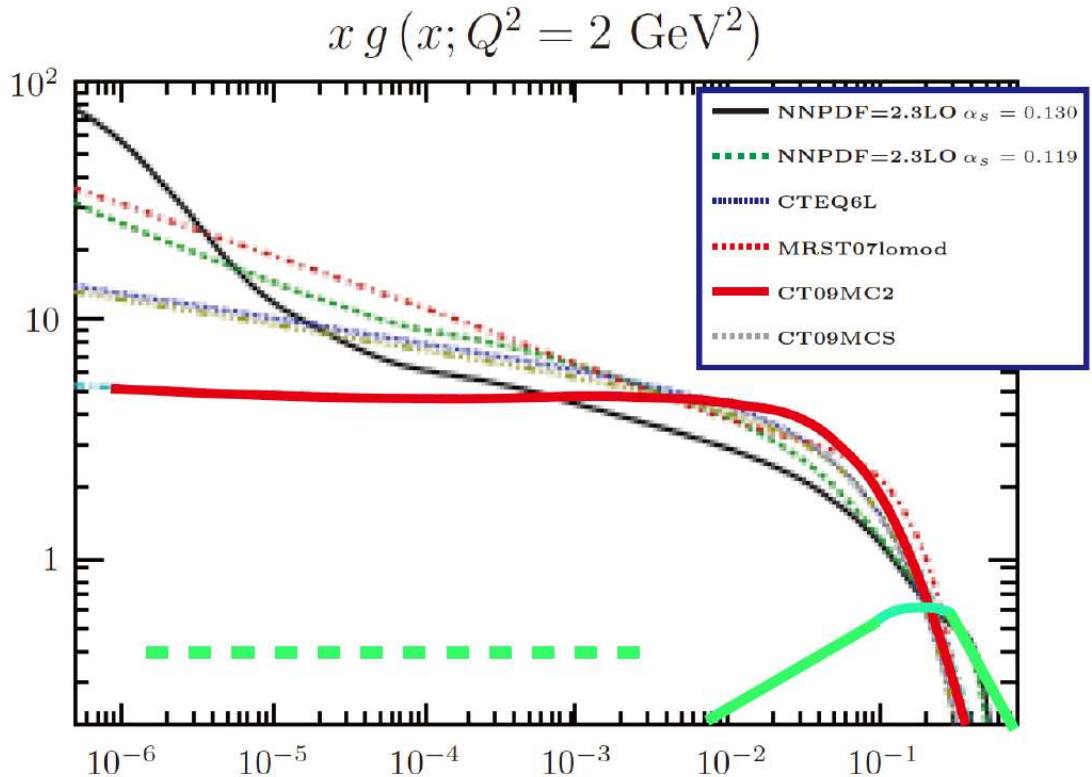}
\end{center}
\caption{The gluon distribution function for several models from Ref. \cite{CMSgluon14}. For the thick red curve for example, it rises from 0 at $x\simeq 0.36$ to 5 at $x\simeq 10^{-2}$. For comparison, we show the  valence $u$-quark structure function $xu_v(x)$  at $Q^2=1.9\,\gev^2$ with the thick green curve of DESY from Fig. 18 of Ref. \cite{HERAst}.  At low $x$, $xu(x)$ will be mostly see quark distributions as sketched with the thick green dash line.} \label{fig:GlueDist}
\end{figure}

The gluon distribution is given in Ref.  \cite{CMSgluon14} which is shown in Fig. \ref{fig:GlueDist}. The thick red curve may be parametrized as
\dis{
G(x)\equiv xg(x)=\left\{\begin{array}{ll}  
-2.7\log\frac{x}{0.36}+\left( \log\frac{x}{0.36}\right)^2,~&\textrm{for }0.005<x<0.36, \\
0,~&\textrm{for } x>0.36.\end{array}\right.
}
For the quark distribution, Ref. \cite{HERAst} from HERA data may be used. Including the sea quarks, the $u$-quark distribution (shown as the thick green curve and dash line) may be   parametrized as
\dis{
U(x)\equiv xu(x)=\left\{\begin{array}{ll} 5.6(x-0.5)+0.4235,~&\textrm{for }x<0.05,\\
-10(x-0.2)^2+0.03(x-0.2)+0.653,~&\textrm{for }0.05<x<0.2,\\
-\log\frac{x}{0.9},~&\textrm{for }0.2<x<0.9,\\
0,~&\textrm{for } x>0.9.\end{array}\right.
}
These will be used for a rough estimate of cross sections. 
The fusion condition is $x_1x_2=m_Q^2/4E^2$ which for $m_Q=1\,\tev$ is $\approx 0.016$ and $0.006$, respectively, at Run-I and Run-II. Folding to the LHC quark and gluon distributions
\dis{
\sigma_{\rm production}&=\int \sigma(x_1,x_2)u(x_1) g(x_2)\delta(x_2=m_Q^2/4E^2x_1)dx_1dx_2 \\
&=\frac{ c_q^2}{ \Lambda^2}\int_{x_{1\,\rm min}}^{0.9} \sigma\left(x_1,\frac{m_Q^2/4E^2}{x_1}\right)u(x_1) \,g\left(\frac{m_Q^2/4E^2}{x_1}\right) dx_1\\
&= \frac{ c_i^2}{ \Lambda^2}\frac{6 m_{q_i}^2 }{\pi    \Gamma^2  }\,  \ln\left(\frac{2}{1-\cos\theta_{\rm min}}\right) \int_{x_{1\,\rm min}}^{0.9}  dx_1\,\frac{U(x_1)G(\frac{m_Q^2/4E^2}{x_1})}{x_1+\frac{m_Q^2/4E^2}{x_1}}\\
&\simeq  \frac{ c_i^2}{ \Lambda^2}\frac{6 m_{q_i}^2 }{\pi    \Gamma^2  }  A^i_{Run}=  \frac{ 1 }{m_Q^2  }\, \frac{  m_{q_i}^2 }{m_Q^2  }\, \frac{3\Lambda^2}{2c_i^2m_Q^2}\,A^i_{Run} ,~\textrm{with }A^u_{Run}\simeq \textrm{  9 and 20},~\textrm{with }A^b_{Run}\simeq \textrm{  0.0215 and 0.26},\label{eq:ExpCS}
}
respectively, for  Run-I ($Run=I$) and Run-II ($Run=II$), because
the rapidity cut at CMS was 2.5, which corresponds to $\theta_{\rm min}=9.38^{\rm o}$. Thus,  $ \ln(2/(1-\cos\theta_{\rm min})) \simeq 5$. For Run-I and Run-II, $x_{1\,\rm min}\simeq 0.016, 0.006$, respectively, and
\dis{
I_R&=\int_{x_{1\,\rm min}}^{0.9} dx_1\,\frac{U(x_1)G(\frac{m_Q^2/4E^2}{x_1})}{x_1+\frac{m_Q^2/4E^2}{x_1}} \simeq 1.80\textrm{ at Run-I, and } 3.92\textrm{ at Run-II},\\
I_R&=\int_{x_{1\,\rm min}}^{0.005} dx_1\,\frac{U(x_1)G(\frac{m_Q^2/4E^2}{x_1})}{x_1+\frac{m_Q^2/4E^2}{x_1}} \simeq 0.0043\textrm{ at Run-I, and } 0.052\textrm{ at Run-II},\label{eq:valence}
}
where $x_{1\,\rm min}$ corresponds to the vanishing gluon distribution above $x_2>0.36$. For $A^i_{Run}$, the sea quark distribution, for example for the $b$-quark, we used only the dash line part with a guessed upper limit of $x_1=0.005$ of Fig.  \ref{fig:GlueDist}. For the $t$-quark sea, the upper bound may be too low to allow any significant number.
Note that the expression (\ref{eq:ExpCS}) does not have $1/E^2$ dependence because of the transition magnetic moment type interaction (\ref{eq:effGlQ}). This feature is helpful in fitting Run-I and Run-II data simultaneously, because we do not have an extra factor $(13/8)^2\simeq 2.64$ for predicting Run-I cross section after fitting to the Run-II data. For example, the CMS data \cite{ATLASCMS,CERNth} are $(0.5\pm 0.6)$ fb at Run-I and  $(6\pm 3)$ fb at Run-II \cite{CERNth}. With the ratio $A^u_{R_I}/A^u_{R_{II}}=0.45$ for the $u$-quark distribution, Run-I data is within 2$\sigma$ level after fitting the Run-II data by Eq. (\ref{eq:ExpCS}). Now, $\sigma_{\rm production}$ given in Eq. (\ref{eq:ExpCS}) crucially depends on the (parton) quark mass and the coupling $c_i$, viz $m_{q_i}^2/c_i^2$. With the ratio $A^b_{R_I}/A^b_{R_{II}}=0.0043/0.052=0.08$ for the $b$-quark distribution, Run-I and Run-II data are simultaneously fitted by Eq. (\ref{eq:ExpCS}).  So, with comparable couplings $c_i$, heavy sea quark contributions dominate. Neglecting the valence quark contributions,  $\sigma_{\rm production}$ is proportional to
\dis{
\sum_i \frac{m_{q_i}^2}{c_i^2}\to \frac{m_b^2}{c_b^2},
}
where in the last equation  the sea bottom quarks are used. For $m_b=4\,\gev, m_Q=1\,\tev$
 \dis{
\sigma_{\rm production}\approx   1.25\times 10^{-40}\,  \frac{\Lambda^2_{\rm TeV}}{c_b^2}\,[\rm cm^2].
 }
 Thus, $\Lambda_{\rm TeV}/|c_b|=O(80)$ will give a reasonable fit.
 
The $Q$ decay BR to $q+P$ is similar to the $Q$ decay BR to $q+\sigma$ because $P$ is the phase field of $\sigma$.
This branching ratio $B_P(Q\to q+P)$ depends on the details of the model. Assuming Eq. (\ref{eq:QtoGlue}) is the leading decay, the probability producing one $P$ from $Q$ decay is 1. As an illustrative example, assume that $P\to 2\gamma$ decay is 10\% without the gluon anomaly coupling. Then, the requirement of $\sigma_{\rm production}\cdot $(branching ratio(BR) to di-photons) at the level of $10^{-38}$ is achieved for  $\Lambda_{\rm TeV}/|c_b|=O(80)$.  If Eq. (\ref{eq:QtoP}) is dominant in the decay of $Q$, a different parameter set should be chosen.

\section{Conclusion}\label{sec:Conclusion}

Axizillas at TeV scale are prospective simple extensions of the SM. In this scheme, we showed that an axizilla produced through the decay  of a TeV scale heavy quark can interpret the di-photon resonance hinted from the LHC Run-II data.  
In this analysis, we assumed the symmetry principle that $P$ is a pseudo-Goldstone boson arising from breaking a global symmetry U(1)$_\Sigma$, and the TeV scale particles are heavy quarks $Q$ and the axizilla $P$.  
   
\acknowledgments{I would like to thank  Ki-Young Choi, Bumseok Kyae, Hyun Min Lee and  Unki Yang for useful discussions. This work is supported in part by the National Research Foundation (NRF) grant funded by the Korean Government (MEST) (NRF-2015R1D1A1A01058449) and by the IBS (IBS-R017-D1-2016-a00). }

\end{document}